\documentstyle[12pt]{article}
\advance\textheight by \topskip
\begin{document}
\begin{flushright}
SU-ITP-96-7\\
hep-th/9602014\\
February 4, 1996\\
\end{flushright}
\vspace{1cm}
\begin{center}
\baselineskip=16pt

{\Large\bf   E(7) SYMMETRIC AREA\\
\vskip 0.6 cm
 OF THE BLACK HOLE HORIZON  }  \\

\vskip 2cm

{\bf Renata
Kallosh\footnote { E-mail:
kallosh@physics.stanford.edu}  ~ and ~\,Barak Kol\footnote{ E-mail:
barak@leland.Stanford.EDU}}\\
 \vskip 0.8cm
Physics Department, Stanford University, Stanford,   CA 94305-4060, USA\\
\vskip .6cm

\vskip 1 cm

\end{center}
\vskip 1 cm
\centerline{\bf ABSTRACT}
\begin{quotation}

Extreme black holes with 1/8 of unbroken $N=8$ supersymmetry are characterized
by the   non-vanishing area of the horizon.  The   central charge matrix
has four generic eigenvalues.  The  area is proportional to   the square root
of  the invariant quartic form of  $E_{7(7)}$. It vanishes in all cases when
1/4 or 1/2 of  supersymmetry is unbroken. The supergravity non-renormalization
theorem for the area of the horizon in $N=8$
case protects the  unique U-duality  invariant.
 \end{quotation}
\newpage

\baselineskip=15pt

$N=8$ supergravity has a hidden symmetry of equations of motion under the group
$E_{7(7)}$ as was discovered  by Cremmer and Julia \cite{CJ}. It was shown by
Duff and Lu  \cite{DuffLu} that  this hidden symmetry has their origin in
supermembrane duality, which in turn implies the string duality by simultaneous
dimensional reduction.
More recently Hull and Townsend \cite{HT} provided some arguments that the
discrete
subgroup of   $E_7$ may be an   exact
symmetry of the string theory. The corresponding discrete subgroup is called
$E_{7}({\bf Z})$ and the symmetry is called U duality. It has been also
emphasized by Witten \cite{W}  that the non-perturbative string dynamics has
its deep origin in 11 dimensional supergravity,
which is known to be the source of the hidden symmetry in 4 dimension.

The purpose of this letter is to show that the area of the horizon of the
extreme black holes with  unbroken supersymmetry can be  studied  from the
perspective of $N=8$ supergravity. It has been understood some time ago that
the supersymmetric bounds on the ADM mass $M$ as well as the charge
quantization are U-duality invariant \cite{HT}.  However if U-duality is indeed
the symmetry of the theory, we may be able to establish the connection of this
symmetry with the area of the extreme black hole horizons. The basic reason to
look for such connection comes from the fact that the canonical geometry of the
black hole does not change under $E_{7(7)}$ transformations. They affect only
the scalars and the vectors of the theory. Thus one may guess that the simple
formula for the area of the extreme black hole horizon may exist  which has the
following properties:

i) In a generic case  when all 4 values of the moduli of the eigenvalues of the
central charge  matrix  $|z_i|$ are different   one has 1/8 of $N=8$
supersymmetry   unbroken,
\begin{equation}
M= |z_1|, \qquad M>  |z_2|, \qquad M
>|z_3|,
  \qquad M> |z_4| , \qquad A \neq 0 \ ,
\end{equation}
\vskip -0.1 cm
\noindent and the area of the horizon $A$ has to be $E_{7(7)}$ symmetric, or,
with an account taken of black hole charge quantization, $E_{7}({\bf Z})$
symmetric.

ii) When all 4 moduli of the eigenvalues of the central charge matrix coincide,
1/2 of $N=8$ supersymmetry is unbroken, and  the area of the horizon $A$
should vanish,
\begin{equation}
M= |z_1|= |z_2|=|z_3|=
  |z_4| , \qquad A=0 \ .
\end{equation}
\vskip -0.1 cm

iii) For the particular case of i)  studied before with 1/4 of $N=4$
supersymmetry unbroken
\begin{equation}
M= |z_1|, \qquad M>  |z_2|, \qquad
|z_3|=
 |z_4| =0 \ ,
\end{equation}
\vskip -0.1 cm
\noindent the $E_{7(7)}$ symmetric formula should reproduce the result
\cite{KLOPP,KO}:
\begin{equation}
A = 4\pi  ( |z_1|^2  - |z_2|^2) \ .
\label{sl2z}\end{equation}

 It should of course vanish when two  moduli of the eigenvalues of the central
charge matrix coincide,
\begin{equation}
M= |z_1|= |z_2|\ ,
  \qquad A=0 \ .
\end{equation}
This example shows that when $|z_1|  =  |z_2|$ and the unbroken supersymmetry
is doubled  the area shrinks to zero.
The corresponding two-dimensional diamond-like picture was presented in Fig. 1
in \cite{KLOPP}. Inside the diamond the black holes are not extreme and do not
have an unbroken supersymmetry. At each edge of the diamond there is one
quarter of supersymmetry unbroken (different part of $N=4$ for each side).
At the vertices of a diamond the unbroken supersymmetry always doubled, since
each vertex has the supersymmetry of each adjoining edge  of the diamond which
enters into a given vertex. The original picture was drawn for the real values
of central charges.
Later on we have found that the same picture appears to be valid upon $SL(2,
{\bf Z})$ rotation: the area as a function of the moduli of two central charges
given in eq. (\ref{sl2z}) is $SL(2, {\bf Z})$ symmetric, see \cite{KO}.

Now we would like to have an analogous picture in terms of the  moduli of the 4
eigenvalues of the central charge  matrix for $N=8$ supersymmetry with various
vertices of coinciding 2 or 4 central charges corresponding to the  shrinking
area
of the black hole horizon. It is rather difficult to visualize this
multi-dimensional figure with vertices describing the pattern of restoring
double and/or  quartic supersymmetry relative to edges.

Fortunately, the hidden symmetry of $N=8$ supergravity helps to find the
solution.
There are actually not so many possibilities to verify: there exists exactly
one quartic
$E_{7(7)}$ invariant which can be build from one\footnote{In case of two
central charges there exists a symplectic invariant for $E_{7(7)}$ which was
already used by Hull and Townsend \cite{HT} to verify that the  quantization of
charges for two dyon black holes is U-duality symmetric.} central charge matrix
 $Z_{AB}$. To support our conjecture about this ``generalized diamond"
function we have
 to show that the area of the black hole horizon is proportional to the square
root of this invariant.

The quartic invariant \cite{CJ} can be represented in the following simple
form\footnote{The   detailed form in which it is presented in \cite{CJ} is the
following:
\begin{eqnarray}
\diamondsuit &=&  Z_{AB} \bar Z^{BC} Z_{CD} \bar Z^{DA} - {\textstyle{1\over
4}} Z_{AB} \bar Z^{AB} Z_{CD} \bar Z^{CD} \nonumber\\
 \nonumber\\
&+& {\textstyle{1 \over 96 }}   \Bigl(\epsilon_{ABCDEFGH} \, \bar Z^{AB} \bar
Z^{CD} \bar Z^{EF} \bar Z^{GH}  + \epsilon^{ABCDEFGH} \,   Z_{AB}   Z_{CD}
Z_{EF}   Z_{GH} \Bigr) \ .
\end{eqnarray}}

\begin{equation}
\diamondsuit =  {\mbox Tr} \Bigl(Z\bar Z \Bigr)^2 -{\textstyle{1\over 4}}
\Bigl({\mbox Tr}\,Z\bar Z\Bigr)^2 + 4  \Bigl({\mbox P\hskip- .1cm f}\; Z +
{\mbox P\hskip- .1cm f} \; \bar Z \,\Bigr)  \
{}.
\label{diamond}\end{equation}

Here
\begin{equation}
Z_{AB} = (q^{ab} + i p_{ab})\, { (\Gamma^{ab} )_A}^B \ ,
\end{equation}
and $ {(\Gamma^{ab} )_A}^B$ are the $SO(8)$ matrices. The 28 electric $q^{ab}$
and 28 magnetic $p_{ab}$ charges are given in terms of the components of
$2\times 28$ vector $ \hat  {\cal  Z} = \hat V {\cal  Z}$. Here $\hat V$ is the
constant value of the
$E_{7}({\bf Z})$-valued field $V$ and ${\cal  Z}$ is the $2\times 28$  vector
of quantized electric and magnetic charges. The Pfaffian ${  {\mbox P\hskip-
.1cm f}}$ of the
antisymmetric complex matrix  $Z_{AB} $ is defined as
${\mbox P\hskip- .1cm f}\; Z \equiv \epsilon^{ABCDEFGH} \,   Z_{AB}   Z_{CD}
Z_{EF}   Z_{GH} $.

The group $E_7$ acts on $Z_{AB}$ as follows:
\begin{eqnarray}
\delta Z_{AB} &=& {\Lambda_A}^C\, Z_{CB} + {\Lambda_B}^C\, Z_{AC} \ , \\
\nonumber \\
\delta Z_{AB} &=& \Sigma_{ABCD}\, \bar Z^{CD} \ ,
\end{eqnarray}
where ${\Lambda_A}^C$ are 63 antihermitian generators of $SU(8)$, and
$\Sigma_{ABCD}$ are totally antisymmetric and  self-dual generators of
$E_7$  orthogonal to $SU(8)$,
\begin{equation}
 \Sigma_{ABCD} =  {\textstyle{1\over 24}} \  \epsilon_{ABCDEFGH}\,
\bar\Sigma^{EFGH} \ .
\end{equation}
Only the discrete subgroup of  $E_{7(7)}$ is compatible with the quantization
condition on dyon black hole charges. The quartic invariant of  $E_{7(7)}$ is
also an U-duality invariant. Thus we satisfy condition i) by construction.

We may check now our condition ii). For example we may consider an $a=\sqrt 3$
extreme black hole, embedded into $N=8$ supergravity. For electrically charged
solution  with  real positive central charges we have
\begin{equation}
 z_1= z_2=z_3=
  z_4  =M .
\end{equation}
The quartic invariant can be calculated using
 $${\mbox Tr} \Bigl(Z\bar Z \Bigr)^2 = 8M^4\ ,  \qquad {\mbox Tr}\,Z\bar Z = 8
M^2\ ,  \qquad    {\mbox P\hskip- .1cm f}\;  Z = {\mbox P\hskip- .1cm f} \;
\bar Z  = M^4.$$

This gives
\begin{equation}
\diamondsuit = 8 M^4 - 16 M^4 + 4 M^4 + 4 M^4 = 0 .
\end{equation}
For pure magnetic case
\begin{equation}
 z_1= z_2=z_3=
  z_4  =i M  ,
\end{equation}
and since the invariant is quartic in central charges, we get the same result:
vanishing area for the solutions with one half of unbroken $N=8$ supersymmetry.

To verify the condition  iii)  we will use the fact that  for this case we may
consider $Z_{12} =z_1$,
$Z_{34} =z_2$, and have   other elements of $Z_{AB}$ vanishing. This leads to
$${\mbox Tr} \Bigl(Z\bar Z \Bigr)^2 = 2\Bigl(|z_1|^4 +|z_2|^4), \qquad {\mbox
Tr}\,Z\bar Z = 2
 \Bigl(|z_1|^2 + |z_2|^2\Bigr), \qquad {\mbox P\hskip- .1cm f}\;  Z = {\mbox
P\hskip- .1cm f}\; \bar Z  =0.$$
In this case
\begin{eqnarray}
\diamondsuit= 2\Bigl(|z_1|^4+ |z_2|^4\Bigr) - {\textstyle{1\over
4}}\Bigl[2\Bigl(|z_1|^2+ |z_2|^2\Bigr)\Bigr]^2 =
\left(|z_1|^2- |z_2|^2\right)^2\ .
\end{eqnarray}

Thus we have learned the the area $A$ of the $SL(2; {\bf Z})$-symmetric
axion-dilaton horizon \cite{KLOPP,KO} in terms of the quartic invariant of
$E_{7}({\bf Z})$ is given by

\begin{equation}
A = 4\pi \sqrt{ |\diamondsuit |}  \ .
\label{usymmetry}\end{equation}

This shows again that the U-duality symmetric formula for the area indeed
covers  previously known solutions. The area is proportional to the square root
of the quartic invariant. This is in a complete agreement with the fact that
$E_{7}({\bf Z})$ has
$SL(2; {\bf Z})$ as a subgroup:
\begin{equation}
 E_{7}({\bf Z}) \supset SL(2; {\bf Z}) \times SO(6,6; {\bf Z})  \  .
\end{equation}

It is interesting to check  the area formula on more general solutions with all
four central charges non-vanishing.  For example, we can consider the
truncation of $N=8$ supergravity to the form describing $N = 4$ supergravity
interacting with  vector multiplets. We consider the action in the form
\cite{R}:
\begin{eqnarray}
S&=& {1\over 16\pi G} \int d^4x \sqrt {-g}\, \Bigl( R-{\textstyle{1\over 2}}
\left [(\partial \eta)^2 +
(\partial \sigma)^2 +(\partial \rho)^2
\right]  \nonumber\\
\nonumber\\
&-& {e^{-\eta}\over 4}  \left [ e^{-\sigma - \rho}  (F_1)^2 +
e^{-\sigma + \rho}  (F_2)^2 + e^{\sigma + \rho}  (F_3)^2 +e^{\sigma - \rho}
(F_4)^2  \right] \Bigr) \ .
\end{eqnarray}
One can use various versions of the known
double dyon  solutions with 1/4 of unbroken supersymmetry of
$N=4$ theory \cite{CY,R}. These solutions are described by two different
central charges. The detailed description of the corresponding 2 central
charges from the heterotic point of view as well as from the point of view of
the Type II theory compactified on $K^3$ is given in \cite{DLR}.
The simplest solution with 1/8 of unbroken $N=8$ supersymmetry  characterized
by
 four different central charges is:
\begin{eqnarray}
ds^2 & = & -e^{2U} dt^2 +
e^{-2U} dx^2,    \hskip 3 cm e^{4U}=  \psi_1 \psi_3 \chi_2\chi_4  \ ,
\nonumber\\
 \nonumber\\
e^{-2\eta}&=&\frac{\psi_1\psi_3}{\chi_2\chi_4}\ , \, \qquad  \qquad
e^{-2\sigma }=\frac{\psi_1\chi_4}{\chi_2\psi_3}\ , \, \qquad \qquad
e^{-2\rho}=\frac{\psi_1\chi_2}{\psi_3\chi_4}\ , \nonumber\\
 \nonumber\\
F_1&=& \pm d\psi_1 \wedge dt \ ,  \quad
 \tilde F_2= \pm d\chi _1 \wedge dt\ , \quad
F_3= \pm d\psi_3 \wedge dt \ ,  \quad  \tilde F_4 = \pm d\chi _4 \wedge dt \ ,
\end{eqnarray}
where
\begin{equation}
\psi_{1} = \left(  1+{|q|_{1} \over r_{1}}\right)^{-1}  , \quad \chi_{2} =
\left( 1+{|p|_{2} \over r_{2}}\right)^{-1}  , \quad  \psi_{3} = \left(
1+{|q|_{3} \over r_{3}}\right)^{-1}   , \quad \chi_{4} = \left( 1+{|p|_{4}
\over r_{4}}\right)^{-1}  ,
\end{equation}
and magnetic potentials correspond to $\tilde{F}_{2/4}=e^{-\eta\pm
(-\sigma+\rho)}F^*_{2/4}$,  where $^*$
denotes the Hodge dual.
Charges in each gauge group could be placed either in various places : $r_1\neq
r_2
\neq r_3 \neq r_4$ or in just one place $r_1=r_2=r_3=r_4$.  The signs of all
charges
could take any values, without correlation between various gauge groups.
 The resulting configuration is characterized by 4 different central charges
(with $4G=1$) :
\begin{eqnarray}
z_1 &=& (q_1 + q_3) + (p_2 +p_4) \ ,\nonumber\\
z_2 &=& (q_1 + q_3) - (p_2 +p_4) \ ,\nonumber\\
z_3 &=& (q_1 - q_3) + (p_2 -p_4) \ ,\nonumber\\
z_4 &=& (q_1 - q_3) - (p_2 -p_4) \ .
\end{eqnarray}
The mass equals to the largest of the moduli of the eigenvalues of the central
charge matrix $M= {\rm max}  |z_i|  , \; i=1,2,3,4$.
The area is proportional to the square root of the absolute value of the
product of electric and magnetic charges
\begin{equation}
A= 4\pi |q_1 p_2 q_3 p_4|^{1/2} \ ,
\label{four}\end{equation}
and in terms of $N=8$ central charges we have found the area to be  equal to
\begin{equation}
A  = 4\pi \left( {  \sum\limits_i z_i^4 -2 \sum\limits_{i>j} z_i^2 z_j^2 + 8
z_1 z_2 z_3 z_4 } \right)^{1/2}\ .
\label{AA}\end{equation}
Again we see that for $z_3 = z_4 =0$ our formula is reduced nicely to
$
A= 4\pi   (|z_1|^2 - |z_2|^2).
$
The crucial check comes here: will the diamond formula (\ref{diamond})
reproduce this expression?
Yes, it does!  One can verify that  eq. (\ref{diamond}) in the case
$Z_{12}=z_1,  \; Z_{34}=z_2, \; Z_{56}=z_3, \; Z_{78}=z_4$ reproduces   eq.
(\ref{AA}).

We would like to make two comments on the black hole solutions in $N = 8$.
The first one is related to the cosmic censorship conjecture. It says that
naked singularities cannot appear as a result of gravitational collapse. This
does not help much for charged stringy black holes since there are no
elementary particles which would carry the corresponding charges, and therefore
these black holes cannot appear as a result of gravitational collapse anyway.
Supersymmetry, which leads to the Bogomolny bound, sometimes implies the
absence of naked singularities \cite{KLOPP}.  However, the link between
supersymmetry and cosmic censorship is not universal, it does not exist, e.g.
for $a = \sqrt 3$ black holes. It is interesting that for $N=8$ supersymmetry
broken down spontaneously all the way to  1/8 of  supersymmetry does play a
role of a cosmic censor. Indeed as long as one keeps away from all the vertices
where unbroken supersymmetry is doubled or quadrupled, the singularities in
canonical geometry are protected by the horizon, exactly as it was observed in
\cite{KLOPP}  in $N=4$ theory.
One may try to develop the idea of Rahmfeld \cite{R}  that the non-singular
black holes
of the Reissner-Nordstrom type  can be build out of the elementary constituents
(for example from four  $a= \sqrt 3$ solutions) which are singular when free.
One may satisfy the condition of  the absence of naked singularities if one
assumes that  four  of these singular constituents,
F-electropole, H-electropole, F-monopole and H-monopole
 may be  confined  inside the non-singular black hole. Each of the elementary
black hole solutions carries the central charges as follows. The first one has
$z_1= z_2= z_3= z_4$, the second one   has $z_1= z_2= -z_3= -z_4$, the third
one
 has $z_1= -z_2= z_3= -z_4$ and finally the last one has $z_1= -z_2= -z_3=
z_4$.
Each of the elementary constituents breaks 1/2 of the $N=8$ supersymmetry,
however, each one breaks a different part of it.
Four of them can be  placed in four different points in space. When  all
charges are placed at one point in space we have a configuration described by
the geometry of the Reissner-Nordstrom type with the singularity  protected by
the horizon, whose area is given by the unique formula
$
A = 4\pi \sqrt {|\diamondsuit |} \ .
$
The unbroken supersymmetry of all four elementary constituents forms only 1/8
of the $N=8$ supersymmetry, which is the maximum common part of the unbroken
supersymmetry of  all four constituents. As long as all four elementary black
holes are at one point, we have a configuration  with the singularity covered
by the horizon.
If  we take  one of the $a=\sqrt 3$ solutions  outside this area, we would have
a naked singularity. Thus, if one really wants to   avoid the violation of the
cosmic censorship (which may be not necessary in application to stringy
solitons) one may conjecture that  $N=8$ supersymmetry has to be broken
spontaneously down to 1/8. The elementary black holes have to be  confined
inside the horizon for this purpose. This picture is consistent with the idea
of  black holes as elementary particles, suggested by Holzhey and  Wilczek
\cite{Holzhey} in the context of $\sqrt 3$ extreme black holes.

The second comment is about  difference between black hole solutions in $N=8$
and $N=4$ theories. If all four supersymmetric positivity bounds of $N=8$
supersymmetry  are respected for the black hole solutions, there are no
non-trivial massless solutions, since the mass
has to be larger than all four eigenvalues of the central charge matrix.
However, in $N=4$ supersymmetry we have only two positivity bounds to respect,
some combination of charges (left-handed in the heterotic theory or some
specific combination in type II string on $K^3$) do not enter the central
charge matrix anymore. Therefore they do not have to vanish simultaneously with
the vanishing ADM mass, and the massless solitons become available \cite{KB}.

It is interesting to note that the quartic invariant in $E_{7(7)}$ was
constructed  by Cartan \cite{CARTAN,CJ} in a form which is different from the
one which was found later by
Cremmer and Julia and which we used here. It is believed that these two forms
are proportional to each other, however,  to the best of our knowledge, no
proof of it is available.  Cartan's quartic form
is
\begin{eqnarray}
J &=&  x^{ij}  y_{jk}  x^{kl}  y_{li} - {\textstyle{1\over 4}} x^{ij}  y_{ij}
x^{kl}  y_{kl} \nonumber\\
 \nonumber\\
&+& {\textstyle{1\over 96 }}   \Bigl(\epsilon^{ijklmnop} \,  y_{ij}  y_{kl}
y_{mn}  y_{op}  + \epsilon_{ijklmnop} \,   x^{ij}   x^{kl}   x^{mn}   x^{op}
\Bigr) \ .
\label{Car}\end{eqnarray}

This alternative quartic form suggests a very nice  interpretation of the fact
that the area is a product of four charges in eq. (\ref{four}). For this
purpose we have to perform the dual transformation from $SU(8)$ version of the
$N=8$ supergravity to $SO(8)$ version.  This type of double analysis  was used
to study extreme black holes in \cite{KLOPP} where we   used in parallel the
$SU(4)$ and the $SO(4)$ version of $N=4$ supergravity. The electric and
magnetic charges in the first theory become either two electric or two magnetic
charges in the other one. In $N=8$ case
this will lead us to reinterpret all four charges as either magnetic or
electric. For example we will get either pure electric solution with $x^{12} =
q_1, \; x^{34} = p_2, \;
x^{56}= q_3, \; x^{56}=p_4$, or pure magnetic one with $y_{12} = q_1, \; y_{34}
= p_2, \;
y_{56}= q_3, \; y_{56}=p_4$. In both cases the area is reproduced by the
Cartan's quartic invariant. In the first pure electric case we have the
contribution only from the
fourth term in $J$, in the pure magnetic case only the third term contributes.
In both cases we get the same result:
\begin{equation}
J= J_{el}= J_{magn}=  4\, q_1 p_2 q_3 p_4 \ .
\label{elmagn}\end{equation}
This makes it plausible that the $E(7)$ symmetric  formula (\ref{usymmetry})
for the area is
in addition proportional to the square root of the Cartan' s quartic invariant.
\begin{equation}
A \sim \sqrt {|J|} \ .
\label{J}\end{equation}
The first two terms do not contribute to (\ref{elmagn}), since we have used a
very simple solution.
However, some more complicated examples of solutions of the heterotic string
theory were found recently by Cveti\v c and Tseytlin \cite{CY} where the area
does not reduce to the product of 4 charges. It would be very interesting to
promote this solution to $N=8$ theory with 1/8 of unbroken supersymmetry and
verify the  area formula for them. Various  black hole  solutions with
different number of unbroken supersymmetry of $N=8$ theory has been studied
also in \cite{CY8}. Khuri and Ort\'{\i}n  \cite{KHO} have recently classified
various supersymmetric embeddings
into $N=8$ theory of the known $a= \sqrt 3,\, 1,\, {1\over \sqrt 3},\, 0 $
extreme black holes.  This study suggests various possibilities to analyse our
area formula.  In particular, the puzzle of the existence of the
non-supersymmetric dyon embedding can be understood from the $E(7)$ point of
view\footnote{The supersymmetric embedding of the $a=0$ black hole in $N=4$
theory  which was performed in \cite{KLOPP} is a particular case of the
solution described above with $2 Q_R = q_1 + q_3, \; 2 P_R = p_2 + p_4, \;
2 Q_L = q_1 - q_3 =0, \; 2 P_L = p_2 -  p_4=0 \ $ and the area $\sim |Q_R
P_R|$.
Thus it also required at least 4 vector fields, from the perspective of $SO(8)$
theory, to be present in the solution.
}.  It is  clear  from (\ref{Car})
why the $E(7)$ symmetry requires at least 4 electric or 4 magnetic charges in
$SO(8)$ version to be non-vanishing to get a non-vanishing area simultaneously
with 1/8 of the unbroken supersymmetry.
This reflects the fact that the on-shell superfields of $N=8$ supergravity are
$E(7)$ symmetric.

So far all checks on the   extreme black holes in $N=8$ theory with 1/8 of
unbroken supersymmetry completely confirm the area formula (\ref{usymmetry}),
(\ref{J}). Hopefully, more elaborated solutions will lead us to all possible
realizations
of the U-duality.

The arguments in favor of the non-renormalization theorem for the extreme black
hole area of the horizon were
presented in \cite{KLOPP}. They were based on the fact that the unbroken
supersymmetry of the bosonic solution of supergravity is equivalent to the
existence of the fermionic isometries in the corresponding superspace.
Therefore the calculation of the on-shell action cannot produce quantum
corrections as long as the corrections come from the
supersymmetric invariants, local or non-local. Due to the presence of the
fermionic isometries all invariants given by the full superspace integrals are
guarantied to vanish due to Berezin's rules of integration over the
anticommuting variables. Apart from possible supersymmetry anomalies (which are
not expected in $N=8$ supergravity) and possible integrals over the
subsupermanifold  (which were studied before and do not seem to challenge the
solutions with unbroken 1/8 supersymmetry), the non-renormalization theorem for
extreme black holes area seems to have a pretty solid basis. In view of this it
is particularly satisfying that this theorem protects a unique quartic
invariant of  $E_{7}({\bf Z})$.

The main  conclusion of this work is the following.  The largest hidden
symmetry of   supergravity with 133 parameters becomes manifest if one   looks
into the structure of the extreme black hole  horizon. 20 years ago P. Ramond
gave a talk  \cite{Ram} with the following title: ``Is there an exceptional
group in your future? $E(7)$ and the travails of the symmetry breaking". This
prediction most certainly worked for extreme supersymmetric black holes.

\vskip 0.4cm

 Stimulating discussions with A. Linde, A. Sen, A. Schwarz, L. Susskind, J.
Rahmfeld, P. Ramond and E. Witten
are gratefully acknowledged. We are  particularly grateful to A. Strominger for
sharing
with us the expectation that the  properties of the  area of the horizon can be
learned even  before  all black hole solutions are found \cite{Andy}.
This work is supported by the  NSF grant PHY-8612280.

\end{document}